\documentclass[aip,jcp,preprint]{revtex4-1}

\def\bx{{\bf x}}
\newcommand{\be}{\begin{equation}}
\newcommand{\ee}{\end{equation}}
\newcommand{\beq}{\begin{eqnarray}}
\newcommand{\eeq}{\end{eqnarray}}
\newcommand{\VEC}[1]{{\bf #1}}

\newcommand{\s}{\hat{\mbox{\boldmath $\sigma$\unboldmath}}}

\usepackage{amsmath}
\usepackage{graphicx}

\begin{document}

\title{Structural transitions in hypersphere fluids: \\
predictions of Kirkwood's approximation}
\author{Jaros\l{}aw Piasecki}
\author{Piotr Szymczak}
\email{Piotr.Szymczak@fuw.edu.pl} 
\affiliation{Institute of Theoretical Physics, Faculty of Physics, University of Warsaw,
 Ho\.za 69, 00-681 Warsaw, Poland} 
\author{John J. Kozak}
\affiliation{DePaul University, 243 South Wabash Avenue, Chicago, Illinois 60604-2301, U.S.A.}

\begin{abstract}
We use an analytic criterion for vanishing of exponential damping of correlations developed previously (Piasecki et al, J. Chem.
Phys., {\bf 133}, 164507, 2010) to determine the threshold volume fractions for structural transitions in hard sphere systems 
in dimensions D=3,4,5 and 6, proceeding from the YBG hierarchy and using the Kirkwood superposition approximation.
 We conclude that the theory does predict phase transitions in qualitative agreement with numerical studies. 
We also derive, within the superposition approximation, the asymptotic form of the analytic condition for occurence of
a structural transition in the $D\to\infty$ limit .

\end{abstract}
\maketitle

\section{Introduction}

The studies of entropic phase transitions in hard hypersphere systems in dimensions four,
 five, and higher, are at present an active field of research.
The possibility of precise quantitative studies appeared owing to the extension of molecular dynamics 
calculations to systems of hard hyperspheres 
\cite{MT1984}. Further development concerning the problem of freezing has been reviewed and discussed in
 \cite{FSL2001} where numerous references can be found.  
The motivation for studying fluids at $D>3$ given in \cite{FSL2001} stresses the fact that the 
knowledge of fluid behavior in different dimensions can be effectively used to construct the 
corresponding density functional theory. Another reason is that in the theory of phase transitions
one can expect important simplifications in dimensions $D\gg 1$. 
In the case of hyperspheres the solution at $D=\infty$ could be used to 
develop a perturbative approach toward lower dimensions.

The fluid to solid transitions in dimensions four, five, and six have been recently analyzed by advanced computations  \cite{SDST2006}, \cite{MFC2009}, \cite{LBW2010}. In paper
\cite{LBW2010} both molecular dynamics and Monte Carlo simulations have been used to study the 
onset of crystallization as reflected in the structure of the radial distribution function.  Of course, the problem 
of primary importance is then the question of packing of hyperspheres.  An interesting observation of geometrical 
frustration in four dimensions reported in  \cite{MFC2009} is here a good illustration. The extension of numerical analysis 
of the fluid-solid transition to even higher dimensions six and seven can be found in \cite{ER2011}.

The analysis of the fluid-crystal interfacial free energy in four, five , and six dimensions performed in 
\cite{MCFC2009} showed that fluid stability  increased with growing dimension. This interesting observation 
permitted to establish a connection with recent theories of jamming behaviour \cite{SWM2008}, \cite{PZ2009}.

The instability of a hard hypersphere fluid with respect to a hypercubic crystal was analyzed in \cite{BR1988}. 
The authors mobilized density functional theory taking advantage of the exact relation for 
dimension $D=1$ and for $D=\infty$ between the singlet density of an inhomogeneous system 
and the two-particle direct correlation function, and obtained via an analysis of bifurcations (see \cite{K1979}) 
an original estimate of the density of closest packing of hypercubic lattices. 

The question of phase structures appearing in hypersphere systems when the spatial dimension $D$ tends to 
infinity  is the object of intensive studies.  A general discussion of "magic dimensions" for which special 
lattice packings appear can be found in \cite{AW2008},\cite{CS1998}. However, it is still by no means clear what kind of 
correlations persist when $D\gg1$.
The possibility of a simplification at $D=\infty$ has been strongly suggested by the study of Mayer series \cite{WRF1987} .
The publications \cite{SDST2006}, \cite{SST2008}, beyond reporting new results, provide a thorough
 description of the present state of the theory in high dimensions. A most interesting guess from 
existing results formulated in  \cite{SST2008} is that in very high dimensions optimal packings of 
hard hyperspheres will be disordered, subject to decorrelation principle.  This challenging hypothesis 
of disorder eventually replacing closed packed crystalline structures became an 
important and fascinating theoretical question (see e.g. chapter 15.4 in  \cite{AW2008}). 

Interpretation of growing wealth of precise numerical data requires a theory. 
Various theoretical approaches used for hyperspheres have been reviewed in \cite{FSL2001} where
the density functional theory, virial expansions, scaled-particle theory, free-volume theory, 
Percus-Yevick and hypernetted chain integral equations are discussed. However, to our knowledge the 
superposition approximation, well known in the theory of liquids (for a critical review see \cite{GZS2004}) 
has not been systematically analyzed up to now in dimensions higher than three. 
The discussion of the content of this theory for $D>3$ is 
our contribution to the current theoretical studies of hard hyperspheres.

Our main object in the present paper is to investigate predictions of Kirkwood's 
superposition approximation concerning the existence
of phase transitions in hypersphere systems. 
The analytic and numerical results for $D=2$, and $D=3$, have been described in 
our previous work \cite{PSK2010}, showing good agreement with numerical studies. 
It turns out that we can apply the methods developed in \cite{PSK2010} to investigate 
higher dimensions as well. In fact, we have at our disposal an analytic criterion
for structural changes in arbitrary $D$. 
Comparison with existing numerical data confirms the correctness of 
predictions as far as the existence of structural transitions is concerned.  So, although it is not known
to what extent Kirkwood's approximation is valid for $D\gg1$, the qualitative predictions
in arbitrarily high dimensions are certainly worth examining. Owing to our method, we are also able to
derive quantitative results, and compare them with numerical data.

Our starting point is the equilibrium YBG hierarchy under Kirkwood's closure.  
In Section II the integral equation for the radial distribution function is derived. In Section III we solve the 
equation by iterations which permits one to determine the  values of packing fractions corresponding to phase 
transitions in dimensions four,five, and six, and compare them with known numerical results. Section III is 
devoted to the discussion of $D\to\infty $ asymptotics. The paper ends with concluding comments. 

\section{Superposition approximation in D dimensions}

The number density $n_{s}$ of $s$-particle configurations in which hard spheres of diameter $\sigma$ occupy space points $(\VEC r_{1}, \VEC r_{2}, ... , \VEC r_{s})$  can be conveniently written in terms of dimensionless positions $\VEC x_{j} = \VEC r_{j}/\sigma$ as 
\be
\label{ny}
 n_{s}(\VEC x_{1}, \VEC x_{2}, ... , \VEC x_{s}) =n^{s} \prod_{a<b}\theta(|\VEC x_{a}-\VEC x_{b}|-1) y_{s}(\VEC x_{1}, \VEC x_{2}, ... , \VEC x_{s})
\ee
where $n$ is the number density of a uniform equilibrium state. The product of unit step functions $\theta(|\VEC x_{a}-\VEC x_{b}|-1)$ represents the excluded volume factor, and $y_{s}$ is the $s$-particle distribution, depending on dimensionless distances $x_{ab}=|\VEC x_{a}-\VEC x_{b}|, \; 1\le a < b \le s$.

The second equilibrium Yvon-Born-Green (YBG) hierarchy equation for hard spheres in  $D\ge 2$ dimensions has the form (see the derivation for $D=2$ in 
\cite{PSK2010})

\be
\label{BGY}
\frac{d}{dx}y_{2}(x) = n\sigma^{D} \int d\s ( \hat{\VEC{x}}\cdot \s )  \theta( |\VEC{x}-\s | - 1) 
 y_{3}(x, 1, |\bx-\s|)
\ee  
Here $\VEC{x} = |\VEC{x}|\hat{\VEC{x}}=x \hat{\VEC{x}}$ denotes the dimensionless relative position of a pair of hard spheres. Distances are measured in the sphere diameter $\sigma$, so $x=1$  describes a contact configuration.  
$\s$ and $\hat{\VEC{x}}$ are unit vectors. The integration spreads over the solid angle 
with
\be
\label{solidangle}
d\s = \sin^{D-2}(\phi_{1})\sin^{D-3}(\phi_{2}) ... \sin (\phi_{D-2})d\phi_{1}d\phi_{2} ... d\phi_{D-1}
\ee
We choose the coordinate system such that 
\[  \hat{\VEC{x}}\cdot \s = \cos\phi_{1} \]

Under the Kirkwood superposition approximation the dimensionless three-particle density $ y_{3}(x, 1, |\bx-\s|)$
factorizes into the product of two-particle distributions, and we get from (\ref{BGY}) a closed nonlinear equation  
\be
\label{KSA}
\frac{d}{dx}y_{2}(x)= n\sigma^{D} \int d\s ( \hat{\VEC{x}}\cdot \s )  \theta( |\VEC{x}-\s | - 1) 
 y_{2}(x)y_{2} (1)y_{2}( |\bx-\s|)
\ee
Introducing a simplified notation $Y(x)\equiv y_{2}(x)$ we find that the correlation function
 $H(x)=Y(x)-1$ satisfies the equation
\be
\label{corr}
\frac{d}{dx}\ln [H(x)+1]= n\sigma^{D}Y(1) \int d\s ( \hat{\VEC{x}}\cdot \s )  \theta( |\VEC{x}-\s | - 1) 
 [ 1 + H( |\bx-\s|) ]
\ee
We denote by $v(1,D)$ the volume of a unit sphere in $D$ dimensions
\be
\label{volume} 
v(1,D)=\frac{\pi^{D/2}}{\Gamma (1 + D/2)}
\ee
Using the formula  
\be
\label{area}
\int_{0}^{\pi}d\phi_{2}\int_{0}^{\pi}d\phi_{3} ... \int_{0}^{\pi}d\phi_{D-2}\int_{0}^{2\pi}d\phi_{D-1}\sin^{D-3}(\phi_{2}) ... \sin (\phi_{D-2})= (D-1)v(1,D-1) 
\ee
we rewrite equation (\ref{corr}) in the form
\be
\label{H}
\frac{d}{dx}\ln [H(x)+1]= \lambda (D) \int_{0}^{\pi} d\phi \cos\phi \sin^{D-2}(\phi) \,\theta[ x-2\cos\phi ] 
 [ 1 + H( \sqrt{x^{2}-2x\cos\phi + 1}) ]
\ee
with
\be
\label{lamb}
\lambda (D) = n\sigma^{D}Y(1)(D-1)v(1,D-1)  
\ee

The right hand side of equation (\ref{H}) can be further simplified. It contains the integral
\be
\label{R1}
  \int_{0}^{\pi} d\phi \cos\phi \sin^{D-2}(\phi) \theta( x-2\cos\phi ) =  -\frac{\theta(2-x)}{D-1}\left[  1 - \frac{x^2}{4} \right]^{(D-1)/2} 
\ee
 and the integral involving the correlation function 
\be
\label{R2}
 \int_{0}^{\pi}d\phi \cos\phi \sin^{D-2}(\phi) \theta( x-2\cos\phi)  H( \sqrt{x^{2}-2x\cos\phi + 1}\; ) =
\ee
\[ \int_{0}^{\pi/2} d\phi \cos\phi \sin^{D-2}(\phi) 
\left[ \theta( x-2\cos\phi)\, H( \sqrt{x^{2}-2x\cos\phi + 1}\; )
- H( \sqrt{x^{2}+2x\cos\phi + 1}\; )  \right] \]

Upon integrating both sides of equation (\ref{corr}) over the interval $(x, \infty)$ we get
\be
\label{int1}
\ln [H(x)+1] = \lambda (D)\left\{ R_{1}(x)  + R_{2}(x) \right\}
\ee
where
\be
\label{ft}
  R_{1}(x) = \frac{1}{D-1}\int_{x}^{\infty}dz \theta(2-z)\left[  1 - \frac{z^2}{4} \right]^{(D-1)/2} = 
 \frac{\theta(2-x)}{D-1}\int_{x}^{2}dz\, \left[  1 - \frac{z^2}{4} \right]^{(D-1)/2}
\ee
and
\[ R_{2}(x) = \]
\[ \int_{x}^{\infty} dz\,\int_{0}^{\pi/2} d\phi \cos\phi \sin^{D-2}(\phi)  [  H( \sqrt{z^{2}+2z\cos\phi + 1}\; )
- \theta( z-2\cos\phi)\,H( \sqrt{z^{2}-2z\cos\phi + 1}\; )  ]  \]
The term $R_{2}$ can be transformed in the following way
\be
\label{transf1}
R_{2}(x) = \int_{x}^{\infty} dz\,\int_{0}^{\pi/2} d\phi \cos\phi \sin^{D-2}(\phi) [H( \sqrt{(z+\cos\phi)^{2} + \sin^{2}\phi}\; )
\ee
\[ - \theta( z-2\cos\phi )\,H( \sqrt{(z-\cos\phi )^{2} + \sin^{2}\phi } \; )\; ]  \]
\[  = \int_{0}^{\pi/2} d\phi \cos\phi \sin^{D-2}(\phi)\left\{  \int_{x+\cos\phi}^{\infty} dz  - 
\int_{x-\cos\phi}^{\infty} dz \,\theta( z-\cos{\phi})\right\}\, H( \sqrt{z^{2} + \sin^{2}(\phi)} \; )  \]
\[ = - \int_{0}^{\pi/2} d\phi \cos\phi \sin^{D-2}(\phi) \int_{x-\cos\phi}^{x+\cos\phi} dz\, \theta( z-\cos{\phi})\, H( \sqrt{z^{2} + \sin^{2}(\phi)}\; )  \]

Introducing the integration variable $\mu = \sin\phi$ we get
\be
\label{transf2}
R_{2}(x) =  - \int_{0}^{1} d\mu \,\mu^{D-2} \int_{-\infty}^{+\infty} dz\, \theta( z^{2}+\mu^{2}-1)\, \theta(\sqrt{1-\mu^{2}} - |z-x|)\, H( \sqrt{z^{2} + \mu^{2}} ) 
\ee

Putting now $s=\sqrt{z^{2} + \mu^{2}}$ leads to the equality
\be
\label{transf3}
R_{2}(x) =  - \int ds\, s H(s) \theta(s-1) \int_{0}^{1}d\mu\, \frac{\mu^{D-2}}{\sqrt{s^{2}-\mu^{2}}}\theta( \sqrt{s^{2}-\mu^{2}}- \frac{x^2 + s^{2} -1}{2x})
\ee
\[ = - \int ds\, s^{D-1} H(s) \theta(s-1) \int_{0}^{1/s}d\nu\, \frac{\nu^{D-2}}{\sqrt{1-\nu^{2}}}\theta\left( \sqrt{1- \nu^{2} } - \frac{x^2 + s^{2} -1}{2sx}\right) \]
One can further simplify this expression by  using the integration variable $w=\sqrt{1- \nu^{2} } $. Indeed, we have
\be
\label{transf3b}
\int_{0}^{1/s}d\nu\, \frac{\nu^{D-2}}{\sqrt{1-\nu^{2}}}\theta\left( \sqrt{1- \nu^{2} } - \frac{x^2 + s^{2} -1}{2sx}\right) 
= \int_{\sqrt{s^{2}-1}/ s}^{1}dw\, (1+w^2)^{(D-3)/2}\theta\left(  w - \frac{x^2 + s^2 -1}{2xs} \right)
\ee
\[ = \int_{(x^2 + s^2 -1)/2xs}^{1}dw\,\theta(1-|x-s|)(1-w^2)^{(D-3)/2}  \]
The last equality follows from the fact that
\[ \frac{(x^2 + s^2 -1)}{2xs} > \frac{\sqrt{s^{2}-1}}{s}   \]

Using this result we eventually find
\be
\label{transf4}
R_{2}(x) =  - \int_{x-1}^{x+1} ds\, s^{D-1} H(s) \theta(s-1)\int_{(x^2 + s^2 -1)/2xs}^{1}dw\,(1-w^2)^{(D-3)/2}
\ee
We can thus write the integral equation (\ref{int1}) for the two-particle correlation function $H(x)$ of $D-$dimensional hypersheres in the form 
\be
\label{iterations}
H(x)={\cal{L}}H(x) = -1 + \exp\{  \lambda (D)[ R_{1}(x)  + R_{2}(x) ] \}
\ee
where
\[   R_{1}(x)  + R_{2}(x)  = \left\{  \frac{2\,\theta(2-x)}{D-1}\, \int_{x/2}^{1}dw\, (1-w^2)^{(D-1)/2} \right.\]
\[ \left. - \int_{x-1}^{x+1} ds\, s^{D-1} H(s) \theta(s-1)\,  \int_{(x^2 + s^2 -1)/2xs}^{1}dw\, (1-w^2)^{(D-3)/2}  \right\} \]

\section{Iterative solution of the integral equation}\label{results}

The integral equation (\ref{iterations}) is solved by a standard Neumann method with  succesive over-relaxation. The iterative solutions are then given by
\begin{equation}
H_n = (1-\alpha)H_{n-1} + \alpha {\cal{L}}(H_{n-1})
\end{equation}
where $\cal{L}$ has been defined in equation (\ref{iterations}).
The relaxation parameter $\alpha$ was taken to be 0.1 (for $D=3,\;4$), and 0.05 (for $D=5,\;6$). Iterations were continued until successive values of $H(x=0)$ differed by less than $\epsilon=10^{-5}$, except in the vicinity of the threshold volume fraction $\phi^*$ (see below), where the convergence was slow and the iterations were discontinued at $\epsilon=10^{-2}$.

\begin{figure}
\center\includegraphics[width=5in]{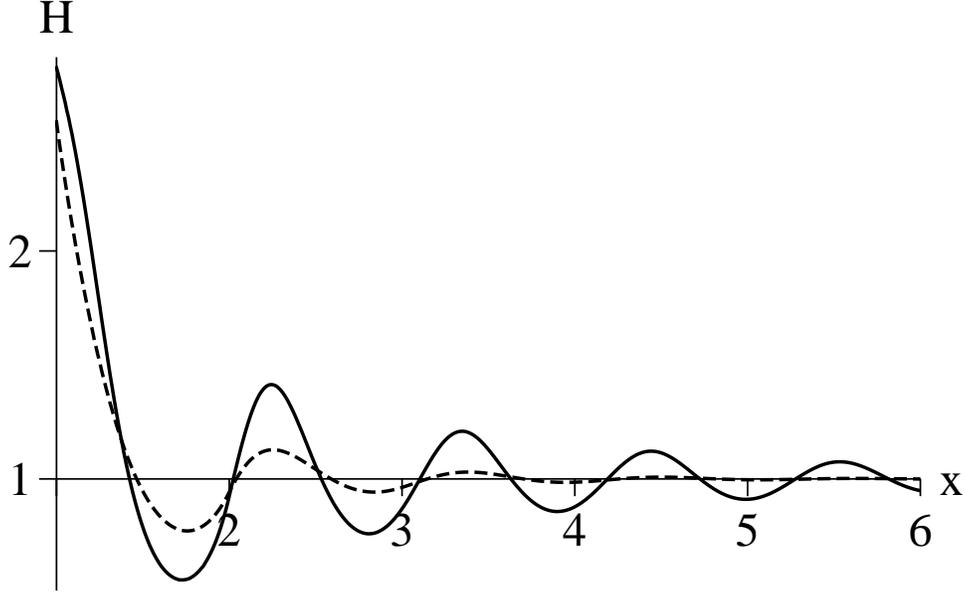}
\caption{Pair correlation function, $H(x)$, for the hard-sphere gas ($D=3$) at the volume fraction $\phi=0.40$ (dashed line) and $\phi=0.51$ (solid line).}
\label{correl3}
\end{figure}

\begin{figure}
\center\includegraphics[width=5in]{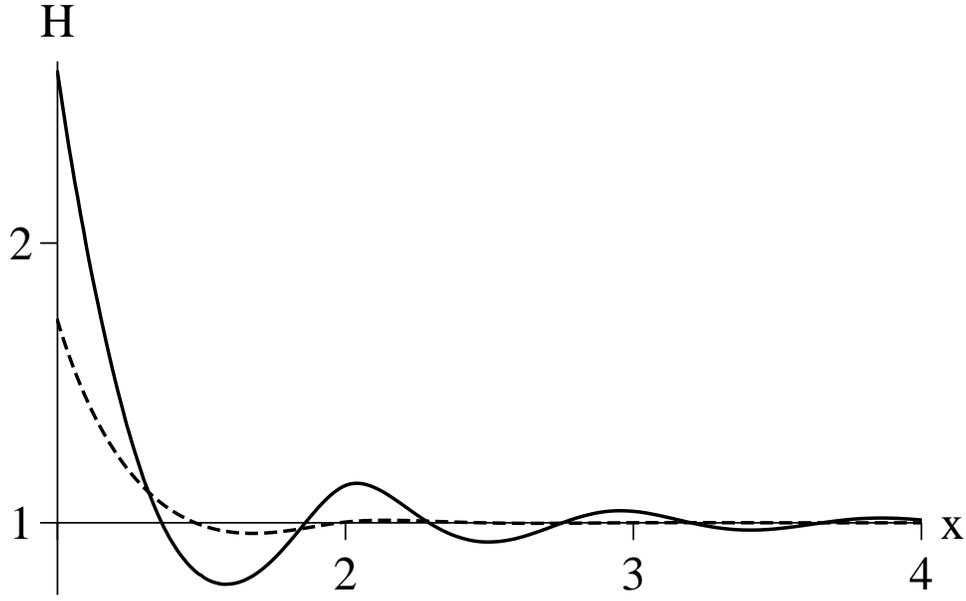}
\caption{Pair correlation function, $H(x)$, for the hard-hypersphere gas ($D=5$) for the volume fraction $\phi=0.10$ (dashed line) and $\phi=0.27$ (solid line) .}\label{correl5}
\end{figure}

Examples of correlation functions obtained in this way are presented in Figs.~\ref{correl3}, ~\ref{correl5}.  
Clearly, with increasing volume fraction the decay of $H(x)$ becomes slower, and a pronounced peak structure appears. 

\begin{figure}
\center\includegraphics[width=5in]{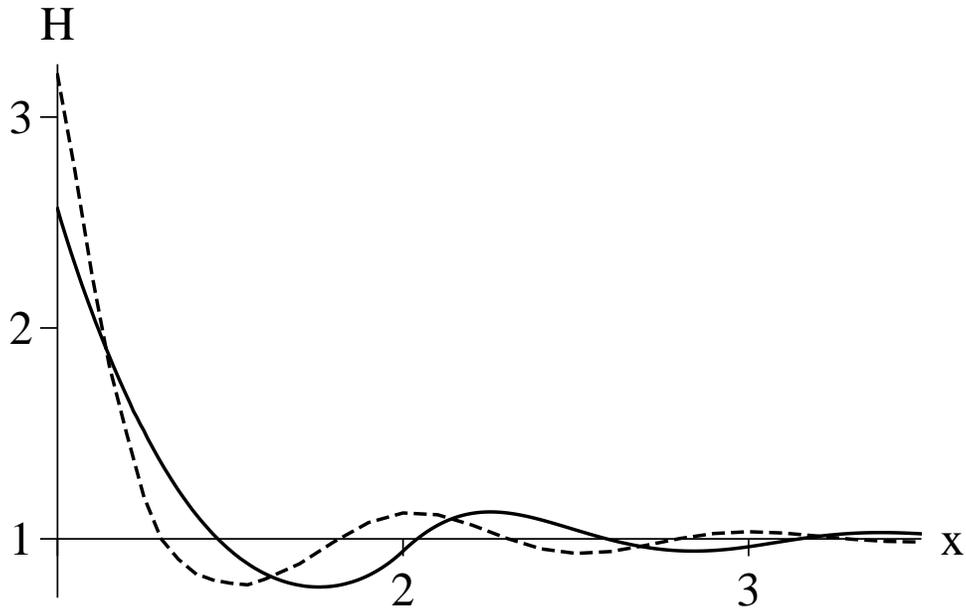}
\caption{Pair correlation function, $H(x)$, for the hard-sphere gas ($D=3$) at the volume fraction $\phi=0.40$
obtained from the solution of the integral equation~\eqref{int1} 
 (solid line) compared with the molecular dynamics simulations of Estrada and Robles~\cite{ER2011} (dashed line).} 
\label{correl3a}
\end{figure}

The comparison of our results with the molecular dynamics simulation data of Estrada and Robles~\cite{ER2011} presented in Fig.~\ref{correl3a} shows that the correlation function obtained from the integral equation \eqref{iterations} has a lower contact value, $H(1)$, and shifted maxima with respect to the molecular dynamics curves. Analogous quantitative differences between the predictions of Kirkwood's approximation and the molecular dynamics data are also observed at higher dimensions and other volume fractions. 


In order to investigate the possibility of a structural change we proceed as in \cite{PSK2010} by considering the form of equation (\ref{H}) for large distances $x$. Using the asymptotic formulae $\ln [H(x)+1] \cong  H(x)$, and $\sqrt{x^{2}-2x\cos(\psi) + 1}\cong x - \cos(\psi)$, we get an integral equation
\be
\label{Hinfty}
\frac{d}{dx}H(x)= \lambda (D) \int_{0}^{\pi} d\psi \cos\psi\,\sin^{D-2}(\psi)  H( x - \cos\psi )
\ee
We then consider the solution $H(x)$ as a linear combination of
exponential modes 
                                \[ H_{\kappa}(x) = \exp( \kappa x ), \]
 where $\kappa$ is a complex number satisfying the equation
\be
\label{kappa}
\kappa =  \lambda (D) \int_{0}^{\pi} d\psi \cos\psi\,\sin^{D-2}(\psi) \exp[ - \kappa\cos\psi  ]
\ee
We look for the mode with the slowest decay.  {\it The disappearance of exponential damping in this mode announces the change in the nature of correlations, and thus implies a structural change.} Such a possibility is equivalent to the appearance of a purely imaginary solution $\kappa =ib$ of (\ref{kappa}), with $b$ obeying
\begin{equation}
\label{immode}
1 + \frac{\sqrt{\pi}}{2}\lambda(D) \Gamma \left(\frac{D-1}{2} \right) \left(\frac{2}{b}\right)^{D/2} J_{D/2}(b)=0
\end{equation}
In deriving equation (\ref{immode}) from (\ref{kappa}) we used the relation
\be
\label{bessel}
\sqrt{\pi}\left(\frac{2}{z}\right)^{\nu}\Gamma(\nu + 1/2)J_{\nu}(z) = \int_{0}^{\pi}\, d\psi \sin^{2\nu}(\psi)\cos(z\cos\psi)
\ee
together with
\be
\label{relation}
\frac{d}{dz}\frac{J_{\nu}(z)}{z^{\nu}} = -\frac{J_{\nu +1}}{z^{\nu}}
\ee
Equation (\ref{immode}) has a solution if and only if $\lambda(D)\ge\lambda^{*}(D)$ where 
\begin{equation}
\label{lastar}
\lambda^{*}(D)
=-\left\{ \frac{\sqrt{\pi}}{2} \Gamma \left(\frac{D-1}{2} \right)2^{D/2} {\rm Min}\left[ \frac{J_{D/2}(b) }{b^{D/2}}\right]\right\}^{-1}
\end{equation}

\begin{figure}
\center\includegraphics[width=5in]{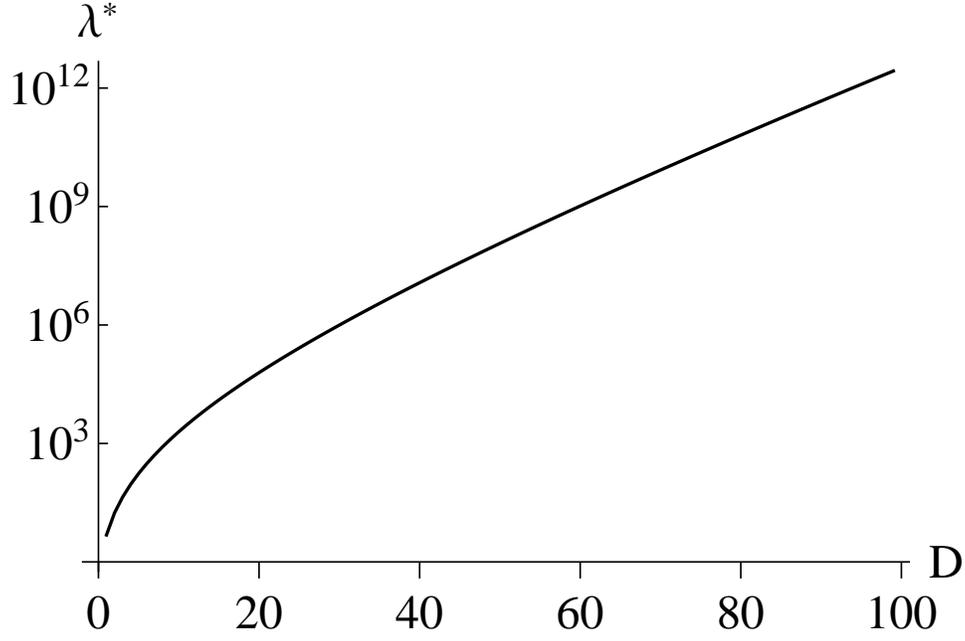}
\caption{Critical value $\lambda^{*}$ as function of $D$.} \label{lstar}
\end{figure}

In order to evaluate the absolute minimum in (\ref{lastar}) we note that according to relation (\ref{relation}) all extrema of function $J_{\nu}(b)/b^{\nu}$ for $b\ne 0$ are attained at points which are zeros of function $J_{\nu +1}(b)$. One finds
\be
\label{bb}
{\rm Min}\left[ \frac{J_{D/2}(b) }{b^{D/2}}\right] =  \frac{J_{D/2}[j(1+D/2, 1)] }{[j(1+D/2, 1)]^{D/2}}
\ee
where $j(1+D/2, 1)$ is the first positive zero of function $J_{1+D/2}$.  Inserting (\ref{bb}) into (\ref{lastar}) we find the following values of $\lambda^*(D)$ :

\begin{itemize}

\item[] $\lambda^*(3) = 17.407$ 

\item[] $\lambda^*(4)= 43.44$ 

\item[] $\lambda^*(5)= 91.23$ 

\item[] $\lambda^*(6)= 172.76$

\end{itemize}
We note here a rapid increase of the critical value $\lambda^*(D)$ with dimension. Indeed, the formula (\ref{lastar}) implies a rapid growth 
(faster than $(e/2)^{D/2}$) illustrated on Fig.~\ref{lstar}.
(In our paper \cite{PSK2010} we found at $D=3$ the threshold value 34.81 for the parameter considered by Kirkwood. et al.\cite{KMA1950}, and equal to
$ 2\lambda$ of the present paper).

\begin{figure}
\center\includegraphics[width=5in]{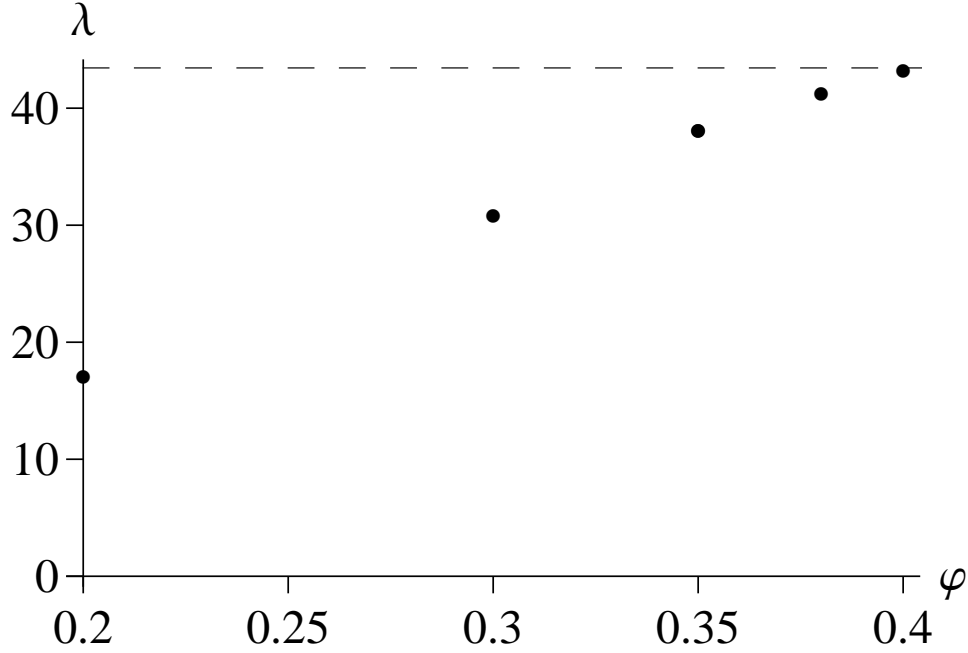}
\caption{The dependence $\lambda(\phi)$ for $D=4$. The dashed line indicates the threshold value, $\lambda^*=43.44$.}\label{lambda}
\end{figure}

The function $ \lambda (D)$ defined in (\ref{lamb}) is related to the volume fraction $\phi(D) $ occupied by the spheres
\be 
\label{vf}
\phi(D) = n\left( \frac{\sigma}{2}\right)^{D} v(1,D)
\ee
by the equation
\be
\label{lv}
\lambda (D) =2^{D}Y(1)(D-1)\frac{v(1,D-1)}{ v(1,D)} \phi(D) 
\ee
where 
\be
\label{ratio}
\frac{v(1,D-1)}{ v(1,D)}=\frac{\Gamma(1+D/2)}{\sqrt{\pi}\Gamma[\,(1+ D)/2\,]} 
\ee
Of course, the contact value of the radial distribution also depends on the volume fraction which we will 
highlight in what follows by the notation $Y(1,\phi)$. 
Example of the resulting $\lambda(\phi)$ dependence (for $D=4$) is given in Fig.~\ref{lambda}. 

Taking $Y(1,\phi)$ from the iteration procedure we can estimate the hyper-volume fractions $\phi^*(D)$ corresponding to the above-derived critical $\lambda^*(D)$. In this way, we get

\begin{itemize}

\item[] $\phi^*(3)=0.52$ 

\item[] $\phi^*(4)=0.40$ 

\item[] $\phi^*(5)=0.28$ 

\item[] $\phi^*(6)=0.21$ 

\end{itemize}

The contact values $Y(1,\phi(D))$ evaluated at $\phi^*(D)$ show with increasing $D$ a decreasing behavior
\begin{itemize}
\item[] $Y(1,\phi^*(3))=2.79 $
\item[]  $Y(1,\phi^*(4))=2.67$
\item[] $ Y(1,\phi^*(5))=2.61$
\item[]   $Y(1,\phi^*(6))=2.51$
\end{itemize}

The Kirkwood approximation does predict phase transitions in dimensions 3, 4, 5, and 6 because 
the threshold values of the volume fractions given above are {\it lower} than the optimal volume fractions
$\phi_{max}(D)$. Indeed, in three dimensions we know the exact result $\phi_{max}(3)= \pi/3\sqrt{2}=0.7404$, 
whereas in dimensions 4 ,5, and 6 the largely accepted conjectures for the densest lattice packings \cite{SDST2006}, \cite{CS1998} yield
$\phi_{max}(4) = \pi^{2}/16 = 0.6168$,  $\phi_{max}(5) = 2\pi^{2}/30\sqrt{2}=0.4652$, $\phi_{max}(6) = \pi^3/48\sqrt{3} = 0.372$.

The numerical results reported in \cite{MCFC2009} show that the phase coexistence region for $D=3$ is in the density range 
$0.494<\phi<0.54$, whereas for $D=4$ and $D=5$ the corresponding ranges are given by
$0.288<\phi<0.337$ and $0.174<\phi<0.206$, respectively. Finally, 
 for $D=6$  the estimated coexistence range corresponds to  $0.105<\phi<0.138$. 
Thus the Kirkwood approximation overestimates the transition point, situating it 
for $D=4,5,$ and $6$ beyond the coexistence region. The discrepancy between our results and the numerical ones measured by the 
ratio $|\phi^*(D) - \phi_{fr}(D) |/\phi_{fr}(D)$, where $\phi_{fr}(D)$ is the volume fraction at freezing, 
increases with growing dimension. Moreover, the numerical simulations of Ref.~\cite{SDST2006} place the volume fraction of maximally random jammed state at $\phi=0.2 \pm 0.01$ for $D=6$, which suggests that the transition at $\phi^*(D=6)=0.21$ might be kinetically inaccessible. 

As far as the origin of the above discrepancies is concerned the following remark can be made. According to a thorough and subtle analysis of  three-particle correlations in hard spheres performed by B. J. Alder \cite{A1964} the superposition approximation gives very good quantitative results provided one extracts the radial distribution directly from the triplet distribution  without using the YBG hierarchy. His important conclusion is that poor quantitative results of the Kirkwood approximation "...are due to an extreme magnification of the error by the integral equations in which it was introduced."  So, it seems possible that the above mentioned discrepancies between our results and numerical predictions are mainly due to the fact that we apply the superposition approximation to the hierarchy equation.  However, the qualitative conclusions may be correct.

At this point another important problem must also be considered. The discussion of our 
results presented so far assumes that the threshold volume fraction $\phi^*(D)$
corresponds to crystallization. But this cannot be really inferred from our approach. 
All we know is that for $\phi(D) > \phi^*(D)$ correlations change their nature, 
and the law of exponentially damped oscillations must be replaced by another one.
The new law must describe long-range, nonintegrable correlations. This last condition is obviously satisfied by 
crystal structures. But it would be also satisfied by states with correlations decaying according to 
nonintegrable power laws. To our knowledge, such power laws  have not been found in jammed or glassy states~\cite{PZ2009,JJC1996}. 
Consequently, the Kirkwood threshold volume fractions $\phi^*(D)$ cannot correspond to their appearance.
We shall come back to this question at the end of the next section. 

\section{ $D\to\infty$: possibility of structural changes}

We now turn to the discussion of the high-dimensional asymptotics $D\to\infty$ having in view the answer to the fundamental question of
existence of phase transitions within Kirkwood's approximation. It follows from the analysis presented so far that correlations change their nature
provided $\lambda(D)>\lambda ^*(D)$.  We have thus to investigate the content of the inequality
\be
\label{inequality}
\lambda(D) > \lambda ^*(D) = -\frac{2^{1-D/2}}{\sqrt{\pi}\Gamma[(D-1)/2]}\left\{\frac{[j(1+D/2, 1)]^{D/2}}{J_{D/2}(j(1+D/2, 1)}\right\}                    
\ee
Using the relation
\be
\label{rel1}
J_{\nu}[j(1+\nu, 1)] = J'_{\nu +1}[j(\nu +1, 1)]
\ee 
together with asymptotic formulae for large $\nu$ (see \cite{W1995})
\be
\label{asympt}
j(\nu ,1) \cong \nu + {\rm const}\,\nu^{1/3},   \;\;\;\;\;\; J'_{\nu}(\nu) \cong - \frac{3^{1/6}\Gamma(2/3)}{2^{1/3}\pi \nu^{2/3} }
\ee
we find 
\be
\label{IN1}
\lambda ^*(D)|_{D\gg1} = \frac{2^{1-D/2}}{\sqrt{\pi}\Gamma[(D-1)/2]}\frac{(1+D/2)^{D/2}2^{1/3}}{3^{1/6}\Gamma(2/3)}(1+D/2)^{2/3}
\ee
In view of relations (\ref{lv}) and (\ref{ratio}) we can rewrite (\ref{IN1}) as
\be
\label{IN2}
2^{D}(D-1)Y(1, \phi^*(D) )\frac{\Gamma(1+D/2)}{\Gamma((1+ D)/2)} 
 \phi^*(D) = 
\ee
\[ \frac{2^{1-D/2}}{\Gamma[(D-1)/2]}\frac{(1+D/2)^{D/2}2^{1/3}}{3^{1/6}\Gamma(2/3)}(1+D/2)^{2/3}  \]
A straightforward calculation yields then the following large $D$ formula for the threshold volume fraction $\phi^*(D)$
\be 
\label{IN3}
\phi^*(D)Y(1, \phi^*(D)) = {\rm const}\left(\frac{D}{2}\right)^{1/6}\left(\frac{e}{2^{3}}\right)^{D/2}
\ee
where \[{\rm const} = 2^{1/3}{\rm e}/[\sqrt{2\pi} 3^{1/6}\Gamma(2/3)\]

In order to check whether the volume fraction $\phi^*(D)$ satisfying  (\ref{IN3}) can be attained one needs a precise knowledge of the upper bound for possible volume fractions
in $D$ dimensions. One also needs the behavior of the contact value $Y(1, \phi(D))$ for $D\to \infty$. 

 According to the Mayer series study  \cite{WRF1987} the hard-hypersphere equation of state at $D=\infty$
has a remarkably simple form 
\be
\label{infty}
p = nk_{B}T [ 1 + \frac{1}{2}n\sigma^{D}v(1, D) ]
\ee
whereas the exact equation  reads
\be
\label{es}
p = nk_{B}T [ 1 + \frac{1}{2}n\sigma^{D}v(1, D) Y(1, \phi(D))]
\ee
It follows that $Y(1, \phi(D))=1$ for $D=\infty$. 
In fact, we have noticed within our approach the decrease of $Y(1, \phi(D))$ with growing dimension when analyzing data for $D=3,4,5\; {\rm and} \; 6$. This observation suggests that $Y(1, \phi(D))$ could monotonously approach 1 when $D\to\infty$. We will thus consider $Y(1, \phi^*(D))$ in (\ref{IN3}) for $D\gg1$ as a number close to 1. The resulting scaling of the volume fraction at phase transition reads

\be
\label{phiscaling}
\phi^*(D) \sim  \left(\frac{D}{2}\right)^{1/6}\left(\frac{e}{2^{3}}\right)^{D/2}
\ee

A straightforward calculation shows that  the Rogers rigorous upper bound for lattice packings \cite{R1958}, \cite{R1964}
\be
\label{Rogers} 
 \phi(D) < \frac{D}{2^{D/2}{\rm e}}  
\ee
is satisfied by scaling  (\ref{phiscaling}) of $\phi^*(D)$. Indeed
\[  \lim_{D\to\infty} \phi^*(D)  \frac{2^{D/2}{\rm e}}{D} = 0   \]
Also the stronger Kabatiansky and Levenshtein bound\cite{KL1978} 
\[  \phi (D) < \frac{1}{2^{0.5990 D}} \]
does not lead to contradiction with (\ref{phiscaling}) for $D\to\infty$.  We thus conclude that scaling (\ref{phiscaling}) 
is compatible with existing upper bounds for crystals. 

The crossing of the threshold volume fraction $\phi^*(D)$ leads to a change in the structure of
hyperspheres reflected by the appearance of long range correlations. 
As we have already remarked at the end of Section 3, this is the reason why we do not expect $\phi^*(D)$ to announce 
the passage to glassy or jammed states. Let us just note that the scaling 
\be
\label{scalingK}
\phi_{K}(D) =  2^{-D}\, D\,\ln D
\ee
of the Kauzmann point $\phi_{K}(D)$ of the thermodynamic glass transition \cite{PZ2009}, \cite{IM2010} 
follows quite a different law compared to that derived for $\phi^*(D)$ in (\ref{phiscaling}). 
We note that \\ $\lim_{D\to\infty} \phi_{K}(D)/ \phi^*(D)=0$ which implies that for sufficiently high $D$ the glass transition would occur at lower volume fraction than the transition at $\phi^*(D)$.

\section{Concluding comments}

Our main object in this paper was the investigation of the possibility of structural transitions in hypersphere systems within Kirkwood's superposition approximation. To this end we employed the simple criterion derived from the equilibrium YBG hierarchy: 
{\it exponential damping of the oscillating pair correlation function $H(x)$ disappears when the dimensionless parameter
 $\lambda(D)=n\sigma^{D}(D-1)Y(1, \phi(D))v(1,D-1) $ attains the threshold value $\lambda^*(D)$ given by  equation (\ref{lastar})}.
For $\lambda(D)>\lambda^*(D)$, the large distance behavior of correlations is necessarily changed. In order to check whether the transition is possible we had to make sure that the threshold volume fraction $\phi^*(D)$ corresponding to $\lambda^*(D)$ was smaller than the maximal possible value $\phi_{max}$.
This was the most difficult point because it required the knowledge of the contact value of the radial distribution $Y(1, \phi^* (D))$, and thus the solution of the integral equation (\ref{iterations}). We performed this program for $D=3,4,5,$ and 6 concluding that the superposition approximation does predict phase transitions for $3\le D \le 6$, in accordance with numerical results, and is thus qualitatively correct.  However, it yields threshold values of the volume fraction higher than those following from numerical studies for crystallization. 

Our investigation of the situation at $D=\infty$ permitted to derive the asymptotic form 
of scaling (\ref{phiscaling})  for the volume fraction at phase transition  showing consistency of the superposition 
approximation with crystallization.  We checked that the known lattice upper bounds for the maximal
volume fractions in $D$ dimensions are not restrictive enough to eliminate the possibility of 
crystallization in arbitrarily high dimension.

An interesting question left open is the limit $\lim_{D\to\infty}Y(1, \phi^* (D))$. 
We noticed in Section III that the contact values $Y(1, \phi^* (D))$ decreased with increasing dimension $D$.  
According to \cite{WRF1987}, the contact value at $D=\infty$ is simply equal to 1. The evaluation of the above limit 
within the superposition approximation would be thus an important test for this theory.

Let us finally stress the fact that although our approach predicts disappearance 
of the fluid structure characterized by exponentially damped correlations, it cannot predict the precise
nature of the new emerging phase.  Clearly further work needs to be done on this point.

\end{document}